\documentclass[prb,twocolumn,superscriptaddress,showpacs,floats,floatfix]{revtex4-1}
\usepackage{graphicx}
\usepackage[table,xcdraw]{xcolor}
\usepackage[american]{babel}
\usepackage{dcolumn}
\usepackage{bm}
\usepackage{multirow}
\usepackage{array}
\usepackage{hyperref}
\usepackage[none]{hyphenat}

\definecolor{gold}{rgb}{1.0, 0.84, 0.0}
\definecolor{rep}{rgb}{0.5, 0.0, 0.5}

\begin{document}

\preprint{APS/123-QED}

% Título y autores
\title{Exploring Three-Atom-Thick Gold Structures as a Benchmark for Atomic-Scale Calibration of Break-Junction Systems}%

\author{J.P. Cuenca}
\affiliation{Departamento de Física Aplicada and Instituto Universitario de Materiales de Alicante (IUMA), Universidad de Alicante, Campus de San Vicente del Raspeig, E-03690 Alicante, Spain.}
\author{T. de Ara}
\affiliation{Departamento de Física Aplicada and Instituto Universitario de Materiales de Alicante (IUMA), Universidad de Alicante, Campus de San Vicente del Raspeig, E-03690 Alicante, Spain.}
\affiliation{Institute of Physics, École Polytechnique Fédérale de Lausanne (EPFL), CH- 1015 Lausanne, Switzerland.}
\author{A. Martinez-Garcia}
\affiliation{Departamento de Física Aplicada and Instituto Universitario de Materiales de Alicante (IUMA), Universidad de Alicante, Campus de San Vicente del Raspeig, E-03690 Alicante, Spain.}
\author{F. Guzman}
\affiliation{Departamento de Física Aplicada and Instituto Universitario de Materiales de Alicante (IUMA), Universidad de Alicante, Campus de San Vicente del Raspeig, E-03690 Alicante, Spain.}
\author{C. Sabater}%
\affiliation{Departamento de Física Aplicada and Instituto Universitario de Materiales de Alicante (IUMA), Universidad de Alicante, Campus de San Vicente del Raspeig, E-03690 Alicante, Spain.}
\email{carlos.sabater@ua.es}

\date{\today}

\begin{abstract}

We present an in-depth study of electronic transport in atomic-sized gold contacts using Break-Junction (BJ) techniques under cryogenic and ambient conditions. Our experimental results, supported by classical molecular dynamics (CMD) simulations and \textit{ab initio} calculations, provide compelling evidence for the formation of three-atom-thick structures in gold nanocontacts under tensile stress. These findings extend previous studies that confirmed the existence of one- and two-atom-thick chains. Beyond identifying these novel atomic configurations, we introduce a fast and robust calibration method for Break-Junction systems, leveraging the characteristic length of these structures to convert piezo-distance into absolute distance in angstroms. Our approach presents a novel and robust method for calibrating atomic distances in atomic conductor systems at both cryogenic and room temperatures. The results also enable the assessment of electrode sharpness, even under ambient conditions.

\end{abstract}

\maketitle

%\tableofcontents

\section{\label{Intro}Introduction}
In the field of molecular and atomic electronics \cite{Cuevasbook}, one of the  challenges is understanding the structure of electrodes at the nanoscale \cite{Agrait1993, Krans93, Agrait2003, Jan2019}. Typically, the most common techniques used to determine the electronic transport  are the Scanning Tunneling Microscope in its break junction configuration (STM-BJ) \cite{Pascual1993} and the Mechanically Controllable Break Junction (MCBJ) \cite{Krans93,SbRuitenbeek,Krans1995}. Both methods are commonly used at low-temperature (4.2 K) and room conditions. However, measuring electronic transport through these techniques is insufficient to fully reveal the atomic structure. Usually, these experiments are supported by CMD simulations \cite{landman1990atomistic} and \textit{ab initio} electronic transport calculations \cite{Diventrabook,Jauhobook, Cuevasbook,Brandbyge2002,Rocha2006,Pauly2008,ANT1,Ferrer2014}. 

Despite significant progress in understanding the geometry of atomic-sized gold contacts and the number of atoms coordinated between them \cite{Yanson1998, Smit2001, UntiedtJ2C, Sabater2012PRL, Sabater13U, Dednam2015, SabaterPRB2016, RoleSab18, Sabater2020chain, Sabater2022PRB}, only atomic structures with one or two atoms of thickness have been identified so far, particularly at low temperatures. While all the studies have shed light on the structure and behavior of one- and two-atom-thick chains \cite{Yanson1998,Sabater2020chain}, contacts thicker than two atoms have not been extensively studied at low temperatures, let alone under ambient conditions. Therefore, our goal is to investigate the existence of three-atom-thick structures and explore their use for calibrating the break-junction (BJ) system across different environments. 

In this manuscript, we have combined  STM-BJ at cryogenic temperature and MCBJ at room conditions experiments, CMD simulations, and electronic transport calculations. Thanks to this combination,  we have identified three-atom-thick structures. Not only has there been progress in techniques to understand the atomic-sized contact geometries, but significant advancements have also been made in calibrating STM-BJ and MCBJ at low temperatures\cite{Untiedt2002}. However, when it comes to developing new methods for calibrating a MCBJ at ambient conditions, progress has been limited. In our case,  we have used the structures composed by three, two and one- atom-thick to calibrate or convert the applied voltage to the piezo system in relative displacement between electrodes. Once our systems are calibrated, we can represent conductance versus calibrated displacement, allowing us to study the slope of these curves historically named traces of conductance. 

In summary, after unmasking the existence of three atom structures, our approach provides a rapid and robust method for calibrating distances, offering new insights into the geometry of atomic contacts at cryogenic temperature and room conditions. Another important aspect of our study is that, once the calibration is completed and after analyzing the slopes of conductance versus relative displacement curves, we can determine how much atomically sharpened  the electrodes are during the dynamic process of rupture.

\section{\label{Methods}Methods and Materials}
\subsection{\label{Model}Electronic transport in BJ techniques}
The most commonly used experimental techniques for measuring electronic transport in atomic-sized contacts\cite{Agrait2003, Cuevasbook,Jan2019} are the Scanning Tunneling Microscope in its break junction configuration (STM-BJ)\cite{Pascual1993,Agrait1993} and the Mechanically Controllable Break Junction (MCBJ)\cite{Krans93,Krans1995}.

\begin{figure}[h!]
 \centering
 \includegraphics[width=0.4\textwidth]{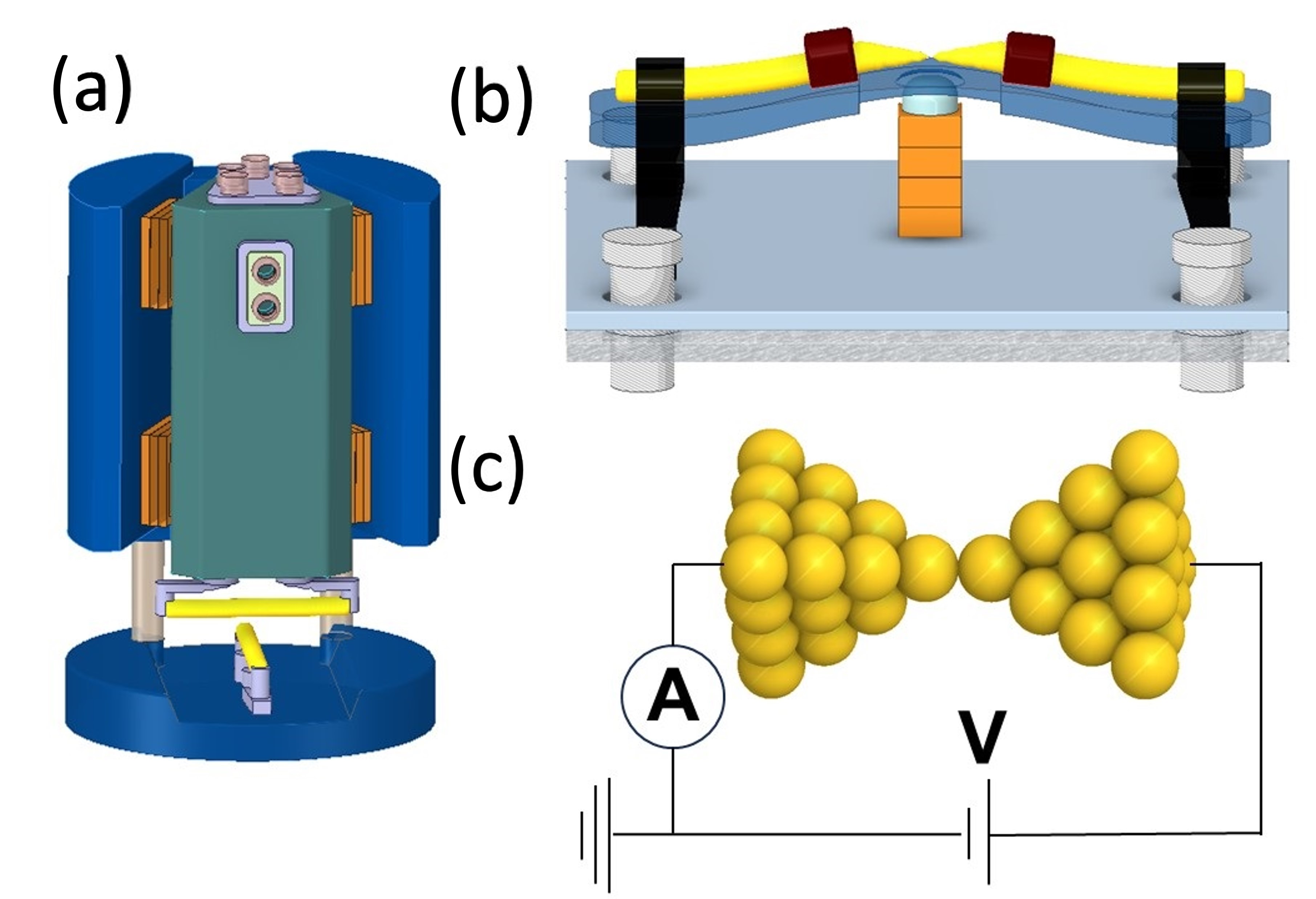}
    \caption{Panel (a) shows an illustration of STM and panel (b) shows the MCBJ experimental setups. Panel (c) illustrates an atomic-sized contact connected in series with a battery and an ammeter.}
    \label{STM_MCBJ}
\end{figure}

In this paper, we analyzed atomic-sized gold contacts using break-junction techniques. Specifically, we performed experiments at low temperatures (4.2 K) using the STM-BJ method, while for experiments under ambient conditions, we employed the MCBJ technique. For the STM, the gold electrodes consisted of two wires with a diameter of 0.25 mm (Goodfellow), positioned facing each other by their circular surfaces and intersecting perpendicularly (see Figure \ref{STM_MCBJ}(a)). In contrast, for the MBCJ, a single notched gold wire was used, grooved, and deposited onto a bendable substrate, as Figure \ref{STM_MCBJ} (b) shows.

Regardless of the method used to create the atomic-sized contact, we measured the conductance using two probes with DC current. Figure \ref{STM_MCBJ} (c) illustrates a schematic of the basic circuit connections for measuring the current in an atomic-sized contact, where the bias voltage ($V_{bias}$) is applied to the atomic contact, which is connected in series with the ammeter. The conductance \( G \) can be expressed in terms of the current \( I \) and the applied voltage \( V_{\text{bias}} \) as \( G = \frac{1}{R} = \frac{I}{V_{\text{bias}}} \). This conductance is often described using the Landauer formalism, which expresses it in terms of the quantum of conductance and the addition of transmission probability \( T_i \) of the \( i \)-th conducting channel, which maximum value per channel can be one. In this framework, the total conductance is given by \( G = G_0 \sum_i T_i \), where \( G_0 \) is the quantum of conductance defined as \( G_0 = \frac{2e^2}{h} \approx \frac{1}{12906}\ \Omega^{-1} \), with \( e \) being the elementary charge, \( h \) Planck’s constant, and the factor of 2 accounting for spin degeneracy (since each quantum channel can carry two spin states). This formulation highlights that conductance at the quantum scale is quantized in units of \( G_0 \), modulated by the transmission probabilities of the available channels.

In our experiments, the bias voltage applied in both configurations (STM-BJ and MCBJ) was 100 mV, and the current was measured by a Femto I/V converter (model DLPCA-200) using an amplification factor of $10^5$   V/A. In our case, the current-to-voltage amplifier is connected to Data Acquisition System (DAQ), which allow us to record the conductance versus relative displacement.

Given that we are working with experiments involving junction rupture, we will represent conductance as a function of the relative distance between electrodes, which can be expressed in Ångströms if the system is calibrated, or in units of voltage applied to the piezoelectric if is not calibrated. Historically, theses curves are called conductance traces. The traces can be either of rupture or formation, depending on whether the nanocontact is being stretched or compressed. In this manuscript, we will focus solely on the analysis of rupture traces.

\subsection{\label{Model} Molecular Dynamics Simulations and \textit{ab initio} calculations}
Classical molecular dynamics (CMD) operates by solving Newton's second law to determine the trajectory of each atom throughout a simulation \cite{frenkel2002,allen1989computer}. In our case, we used CMD to simulate the rupture and formation of a atomic gold \cite
{landman1990atomistic}. The simulation was performed using the \texttt{LAMMPS} \cite{plimpton1995fast, lammps2, LAMMPS3} code with an embedded-atom model (EAM) \cite{Baskes} with the interatomic potential described by Zhou \textit{et al.} \cite{zhou2001atomic,wadley2001potential}. We simulated the pull-push process by applying opposite velocities of $\pm 0.04$ $ \text{Å/ps}$ at each step to the upper and lower layers of the electrode, with steps of 1 ps.  To ensure the formation and rupture of the junction, the rupture and formation processes took 250 ns and 254 ns, respectively. The temperature of the system was maintained at 300 K in an NVT canonical ensemble with a Nose-Hoover thermostat. The input configuration chosen is a gold nanowire oriented along the crystallographic direction (001) with a narrower middle section \cite{RoleSab18}.

Every 5 ns in the simulation, a snapshot of the position was taken, allowing us to prepare the system for the \textit{ab initio} calculations. For that purpose, we reduced the number of atoms in the snapshot by taking only $\sim$ 80 atoms in the vicinity of the minimal cross-section. With a small CMD minimization process target at the outermost layers of the snapshots, we added two layers of gold (001) at the extremes needed for the following transport computation.

To compute the electronic transport over the snapshot, we used the non-equilibrium Green’s function (NEGF) approach. The code used was the Atomistic Nano Transport (\texttt{ANT.G}) code \cite{ANT1, ANT_2,ANT_3, ANT_4, ANT_code}, which interfaces with \texttt{Gaussian} \cite{GAUSSIAN09}. The added gold layers use the CRENBS basis set \cite{Crenbs}, the rest of the atoms have the basis pob-TZVP-rev2 \cite{pob-TZVP-rev2,Stutt_pseudo}, prepared for metallic gold computations. Finally, the DFT computation was done at the GGA level with the implementation of PBE built-in Gaussian \cite{PBE1, PBE2}.

\section{\label{Results}Results and discussion}
\subsection{Low Temperature STM-BJ }
As previously mentioned, the traces represent conductance versus the relative displacement of the electrodes, measured in distance units if the system is calibrated, or in voltage units if the piezoelectric system is not calibrated, regardless of the BJ technique used.

As noted by Untiedt \textit{et al}. in the manuscript \cite{Untiedt2002}, before 2002, a controversy arose regarding the measurement of interatomic distances in gold chains, with some authors reporting a distance of 3.6 Å, while others found 2.5 Å. Thanks to Untiedt’s manuscript, this issue was resolved using three different calibration methods. Two of these methods are based on measurements of current in tunneling barriers, while the third utilizes an STM-BJ system combined with an interferometric setup and length histograms. However, the first two methods require knowledge of the material's work function, which is known to change in non-vacuum environments, such as under ambient conditions. Therefore, these methods cannot be applied in ambient conditions, making the only viable option the measurement of atomic distances through length histograms.  Therefore, one of the objectives of this manuscript is to contribute a new calibration method and reveal the corresponding geometry of structures with a thickness of three atoms.

\begin{figure}[h]
 \centering
    \includegraphics[width=0.49\textwidth]{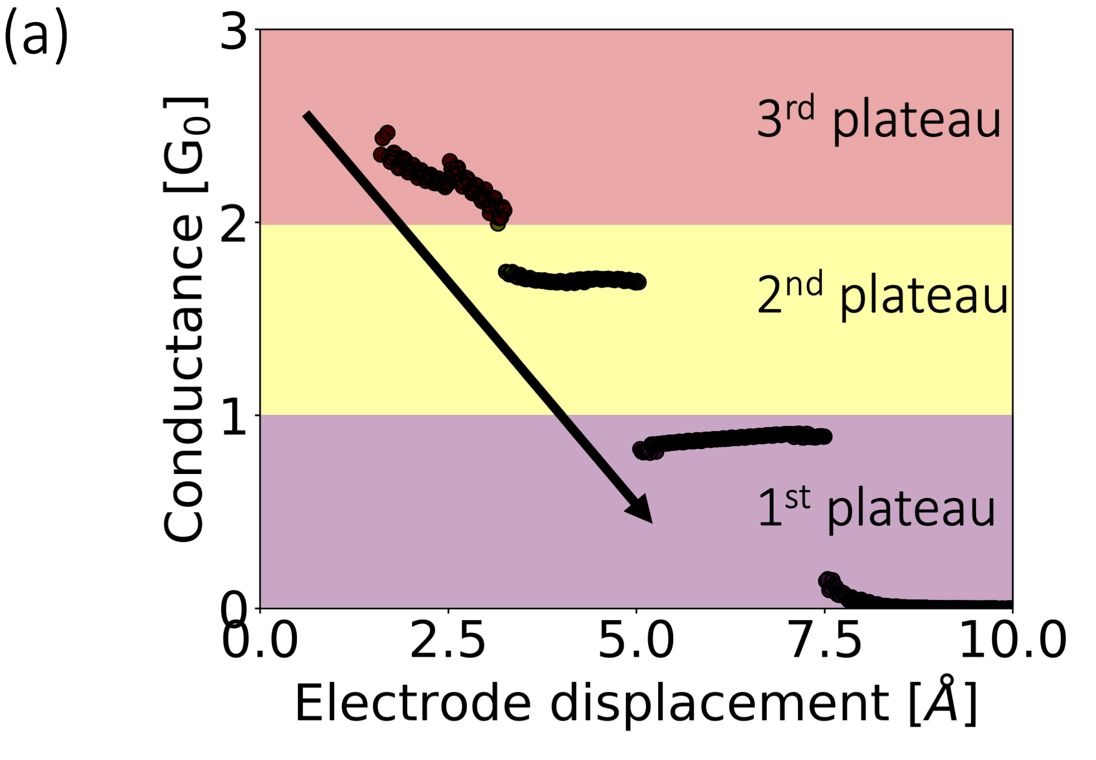}
    \par\bigskip
    \includegraphics[width=0.47\textwidth]{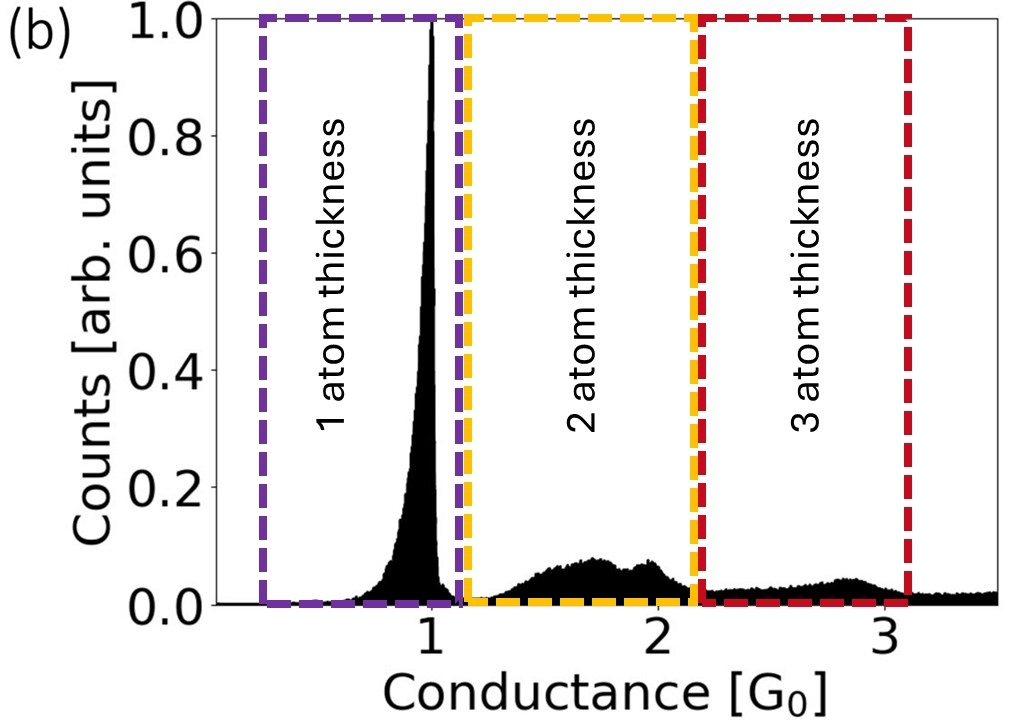}
\caption{Panel (a) illustrates a rupture trace featuring three distinct plateaus, each highlighted in a different color. The purple, yellow, and red regions correspond to the first, second, and third plateaus, respectively, representing atomic-sized geometries with thicknesses of one, two, and three atoms. Panel (b) shows a normalized conductance histogram constructed from 6000 rupture traces, using the same color code to indicate these atomic structures.}
\label{Trace4K}
\end{figure}

To calibrate the distance in our STM-BJ experiments conducted at low temperature (4.2 K), we used the length histogram method, converting the voltage applied to the piezoelectric system into Ångströms (Å) \cite{Yanson1998, Smit2001, Untiedt2002}. Figure \ref{Trace4K}(a) shows a rupture trace of gold at 4.2 K, where the relative electrode displacement is expressed in Å.  In other words, we define a plateau as a region where the conductance remains stable, without abrupt jumps between consecutive points. Three conductance plateaus can be observed in Figure \ref{Trace4K}(a). To facilitate visualization, we assigned a color code—purple, yellow, and red—to indicate the conductance ranges of the first, second, and third plateaus, respectively. These plateaus correspond to cross-sectional structures of one, two, and three atoms in thickness, as also referenced in the following panel and throughout the rest of the paper.

One of the most common statistical analyses performed on conductance traces is the construction of conductance histograms, as they provide insight into the most frequently observed conductance values. These values relate to certain atomic geometries, which appear more often than others during the stretching process.  Figure \ref{Trace4K}(b) shows a conductance histogram of Au traces at 4.2 K in high vacuum. The same color code is applied to the dashed rectangles to help in identifying the three different thicknesses of the structures, corresponding to the previously defined plateaus. This is typically  the characteristic histogram for gold with a distinctive peak at 1 $G_0$,  which has been attributed to the contribution of monomers, dimers, and atomic chains \cite{Yanson1998, Ohnishi1998, Smit2003, Sabater13U, Sabater2015, Otal2014, RoleSab18} (one-atom thickness). Furthermore, the atomic structures associated with the second plateau were revealed in Ref. \cite{Sabater2020chain}(two-atom thickness). However, the nature of the structures corresponding to the third plateau (or three-atom thickness) remains uncertain, with values ranging between approximately 2.1 and 3 $G_0$.

As mentioned earlier, one way to convert piezo voltage to relative distance is through length histograms. In the literature, we can find length histograms already in distance units for both the first \cite{Yanson1998, Untiedt2002, Smit2003, Otal2014, Sabater2015} and second plateau \cite{Sabater2020chain} (also recognized as structures of one and two atoms in thickness). In this manuscript, we performed the length histogram on the plateau that hypothetically belongs to contacts with a thickness of three atoms (third plateau).

\begin{figure}[h!]
 \centering
 \includegraphics[width=0.47\textwidth]{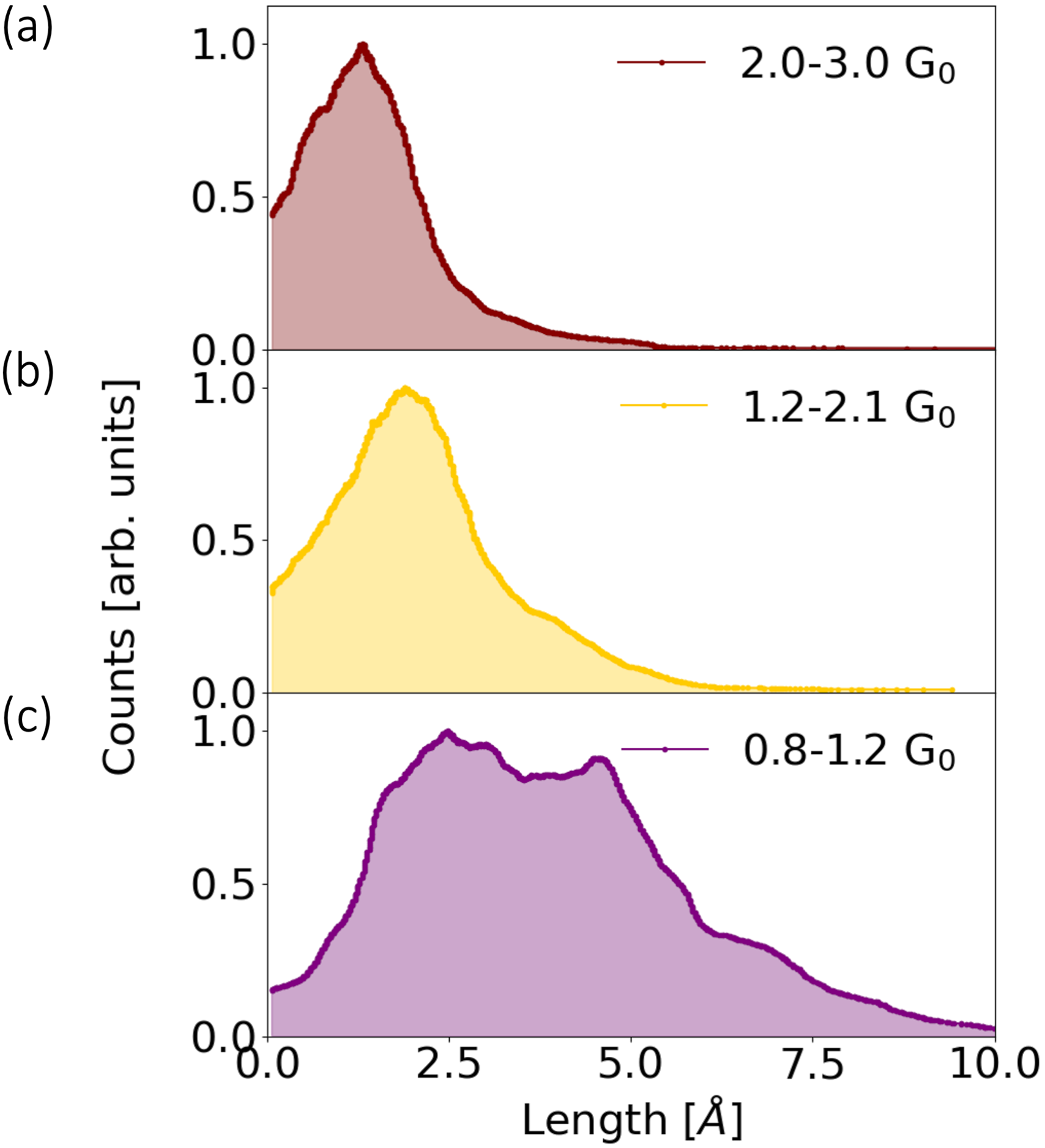}
    \caption{Histogram of lengths for the last three conductance plateaus obtained from rupture traces at 4.2 K in cryogenic vacuum. The bottom panel (c)  corresponds to the conductance range of 0.8–1.2 \(G_{0}\) (purple histogram), the middle panel (b) shows the length histogram for plateaus in the range of 1.2–2.1 \(G_{0}\) (yellow histogram), and the upper panel (a) presents the length histogram for plateaus within the range of 2.0–3.0 \(G_{0}\) (red histogram).}
    \label{Histolength}
\end{figure}

In Figure \ref{Histolength}, we show the length histograms performed on the first, second, and third plateaus in approximately 6000  gold rupture traces at 4.2 K. The color codes purple, yellow, and red represent, respectively, the structures of one, two, and three atoms in thickness. As seen in Fig. \ref{Histolength}, the bottom panel (c) shows the length histogram (purple) of the first plateau characteristic of gold, which is similar to that described in the literature \cite{Yanson1998,Untiedt2002,Smit2003,Otal2014,Sabater2015}. In the middle panel (b), the length histogram (yellow) is presented, which also agrees with previous results described in reference \cite{Sabater2020chain}. Finally, the upper panel (a) histogram (red) represents the length plateaus between 2.1 $G_0$ and 3.0 $G_0$, and as observed, the most typical value is 2.5 \text{Å},  indicating that typically during the breaking process, contacts of approximately one atomic distance and a thickness of three atoms are formed. Moreover, observing the third-length histograms, the maximum frequency peak is 2.5 \text{Å}. That is, regardless of whether we have a structure of one, two, or three atoms in thickness, the distance that is reproduced the most frequently is that of an atomic distance of 2.5 \text{Å}.

\begin{figure}[h!]
 \centering
\includegraphics[width=0.425\textwidth]{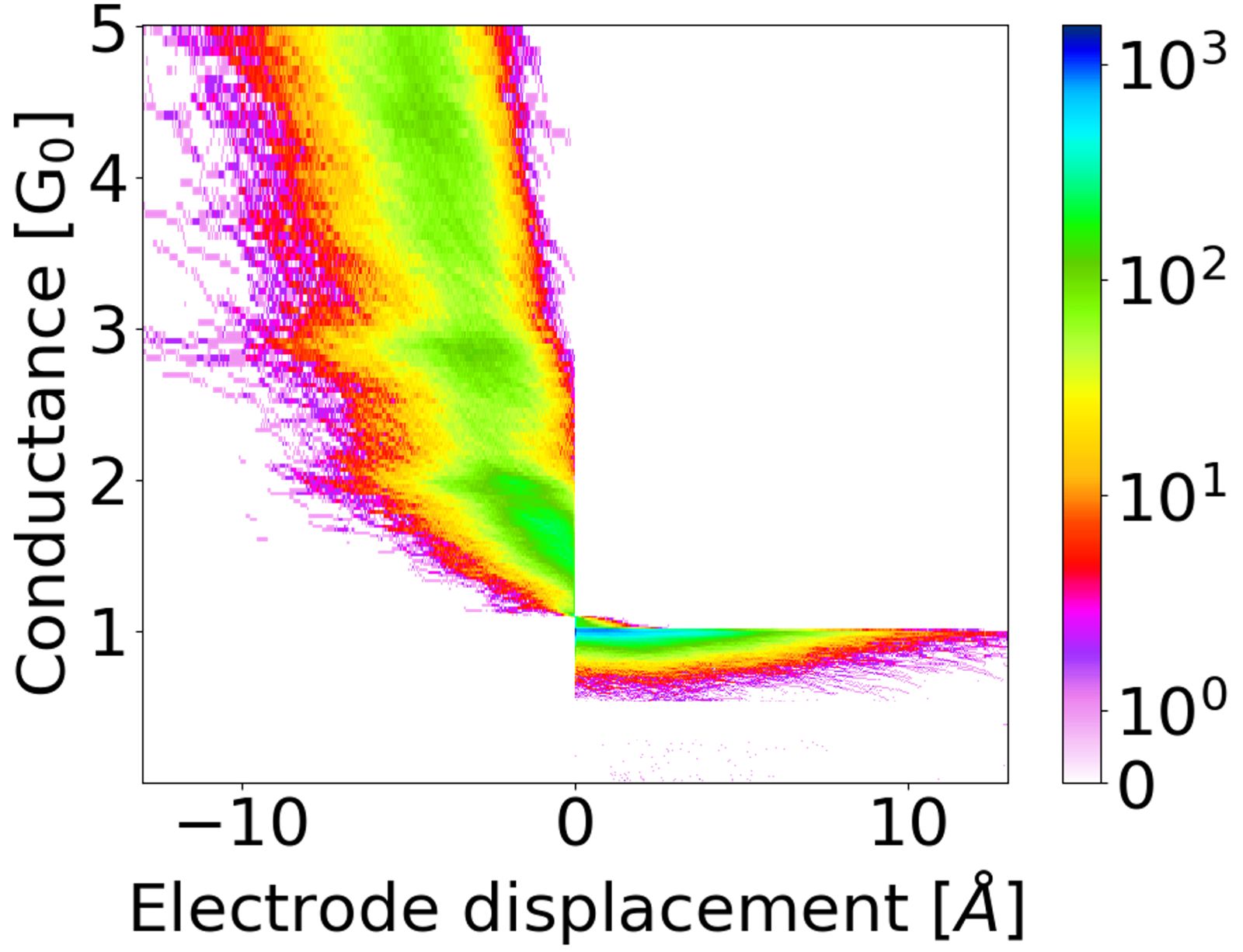}
    \caption{The density plot was constructed from 6000 rupture traces of gold measured at 4.2 K in a cryogenic vacuum. The color bar is on a logarithmic scale, transitioning from warm to cold colors.}
    \label{Densdplot}
\end{figure}

Fig. \ref{Densdplot} shows a density plot. The relative displacement of all traces has been shifted to align at the collapse point, which corresponds to the first value of the first plateau at 
$1 G_0$, when reading the trace from left to right. The color bar ranges from zero counts (white) to ten thousand counts (blue). The green color indicates that the event occurs approximately a hundred times. Fig. \ref{Densdplot} shows that in the range of 0.8–1.0 $G_0$, the green area extends up to 5 Å, which is in clear agreement with the length histogram (red) in Fig. \ref{Histolength} \cite{Otal2014}. Furthermore, the blue-cyan region around $1 G_0$  extends from 0 to 2.5 Å, corresponding to the first peak of the length histogram for the first plateau (purple) in Fig. \ref{Histolength}.

Nevertheless, in a density plot of traces at room temperature, it appears that the point clouds corresponding to the regions of 2.5$G_0$, 2$G_0$, and 1$G_0$  have similar lengths. This suggests that, although they are not calibrated, the most frequent regions in the plot exhibit the same apparent length, even though they are represented in volts rather than length units. Building on this observation, we proposed a novel calibration method for BJ techniques. Specifically, we constructed density plots in units of conductance versus relative displacement. Then, knowing that the regions corresponding to structures of 1, 2, and 3 atoms in thickness correspond to 2.5 Å, as observed at low temperatures, we performed an equivalence calibration along with an estimation of its calibration error.

\subsection{MCBJ Approach at Room Condition}
\begin{figure}[h!]
 \centering
  
 \includegraphics[width=0.49\textwidth]{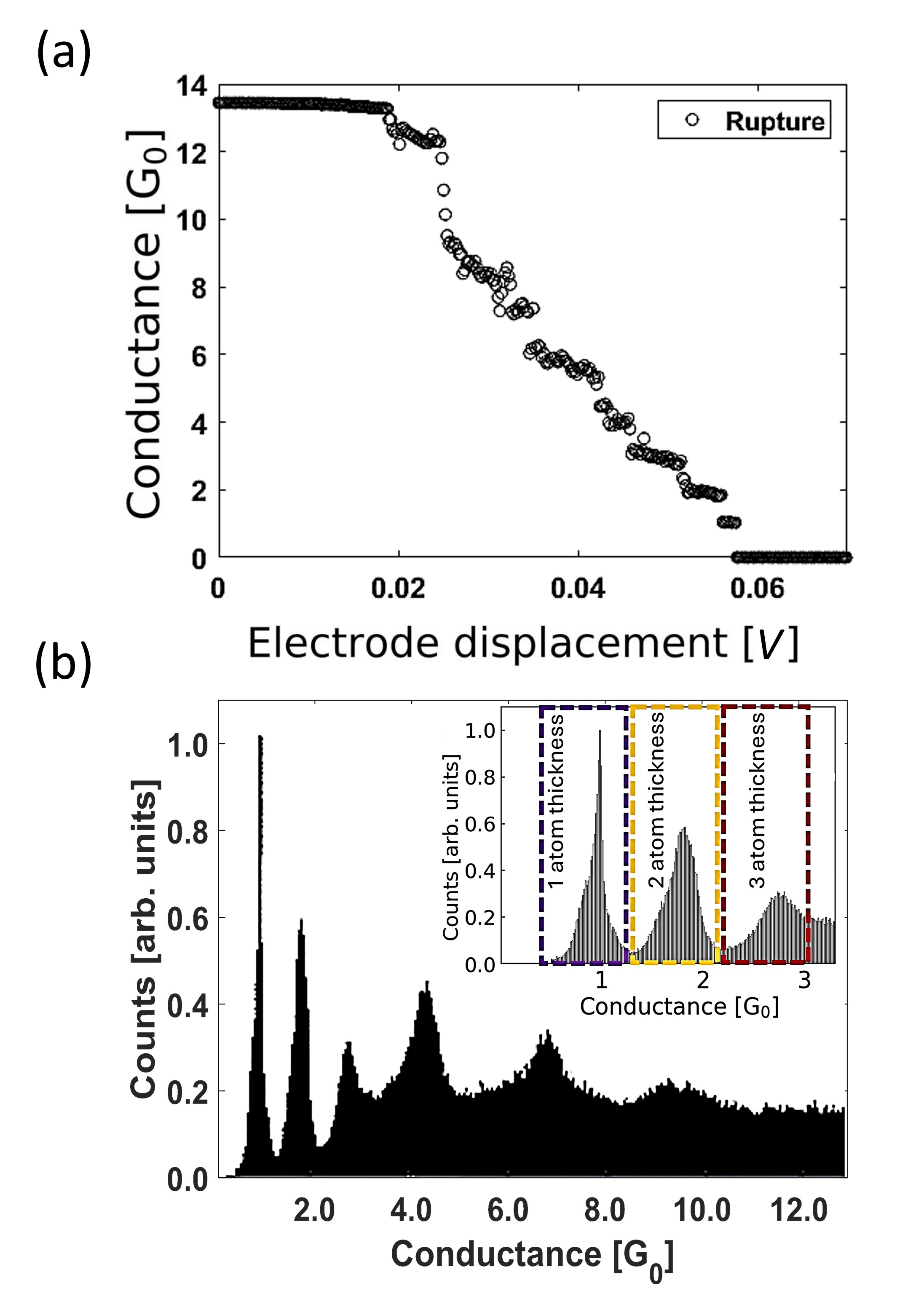}
    \caption{(a) Trace of conductance of gold at room conditions measure in MCBJ. Where the relative displacement is volts units. The arrow indicates the direction to read the trace. (b) histogram of conductance in the range of 0.1 to 12 $G_0$, inset show a zoom in of the range 0.1 to 3 $G_0$, the colored rectangles indicates the structure attributed.}
    \label{NoCaltrace}
\end{figure}

As it is well known, an MCBJ consists of a piezo element responsible for bending the flexible substrate on which the notched wire is mounted (as Fig.\ref{STM_MCBJ} (b) shows). When the piezo pushes against the substrate, it breaks the contact, allowing the measurement of a rupture conductance trace, where the units of relative displacement are in volts, as shown in Fig. \ref{NoCaltrace}(a).

Regardless of whether the electrode displacement is calibrated or not, a conductance histogram can be obtained. Figure \ref{NoCaltrace}(b) presents a typical conductance histogram, normalized in the range of 0.1 to 13 $G_{0}$, constructed from 5000 rupture traces. The inset shows the normalized histogram in the range of 0.1 to 3.2 $G_{0}$, where the highlighted rectangles indicate the conductance ranges attributed to structures of one-, two-, and three-atom thickness. These structures will be revealed through CMD simulations, and their electronic transport properties will be obtained by DFT calculation.

\subsection{Identifying the structure of three atom thickness via MD and DFT}
\begin{figure}[h!]
 \centering
 \includegraphics[width=0.49\textwidth]{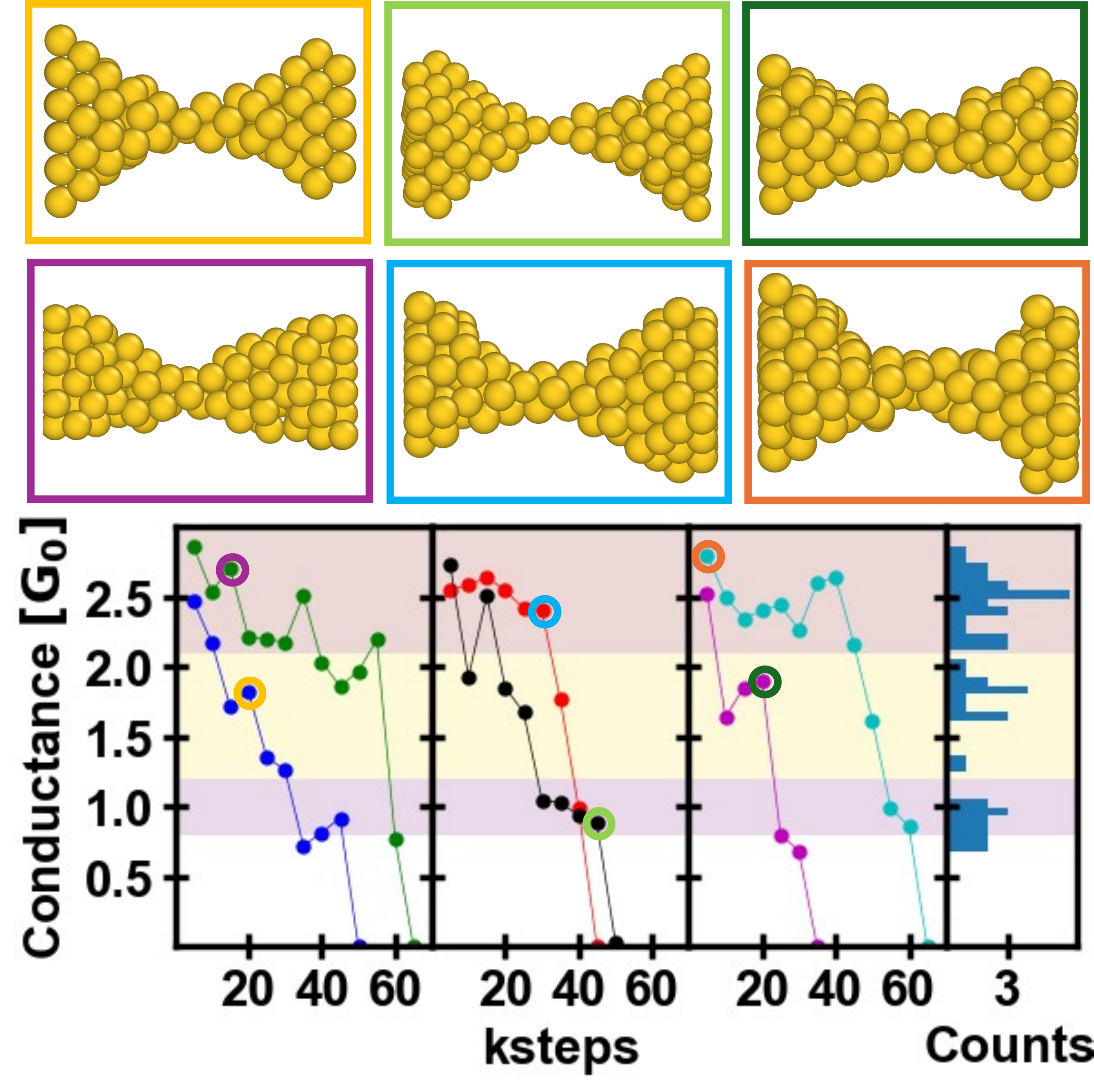}
       \caption{Conductance of the MD snapshots, each color in the graph representing consecutive snapshots. On the right side there is a histogram of the conductance of the calculations. The color of the background is for the conductance range found in different plateaus: (0.8-1.2), (1.2-2.1) and (2.1-3.1) $G_0$. Above the conductance graph, there are schematic representations of the points highlighted in the same color of the frame.}
    \label{FigTeo}
\end{figure}

In Figure \ref{FigTeo}, the upper section shows some of the structures obtained through molecular dynamics simulations of continuous rupture cycles of the gold nanowire. These structures are highlighted with a colored rectangle, where the color code corresponds to the theoretically obtained conductance values.

The lower panel represents the conductance obtained throughout the simulation steps. This panel consists of three subgraphs, each displaying a simulated rupture trace. In each trace, a point marked with a circle indicates the corresponding structure, with a color code matching that of the upper panel. Additionally, in the background of the lower panel, three color-coded regions highlight the zones corresponding to one-, two-, and three-atom thicknesses, as consistently shown throughout the manuscript.

As observed in the figure, DFT calculations have revealed three-atom-thick structures with a triangular packing, exhibiting a conductance of approximately 2.5 
$G_0$. On the other hand, two-atom-thick structures show a conductance close to 1.6$G_0$.  Moreover, single-atom-thick structures present a conductance of approximately 1$G_0$. Finally, the conductance traces show that rupture occurs when the conductance approaches 0 $G_0$.

Lastly, the lower right panel displays a  histogram of the calculated conductances, which shows a resemblance to the experimental ones, as seen in the inset \ref{NoCaltrace} (b). Despite the fact that the histogram is composed only of six hand-picked simulated traces, while the experimental one is composed of thousands of traces, the overall trend remains comparable.

\subsection{New methodology to calibrate a BJ system at room conditions based on experimental, simulation, and calculation results}

\begin{figure}[h!]
 \centering
 \includegraphics[width=0.49\textwidth]{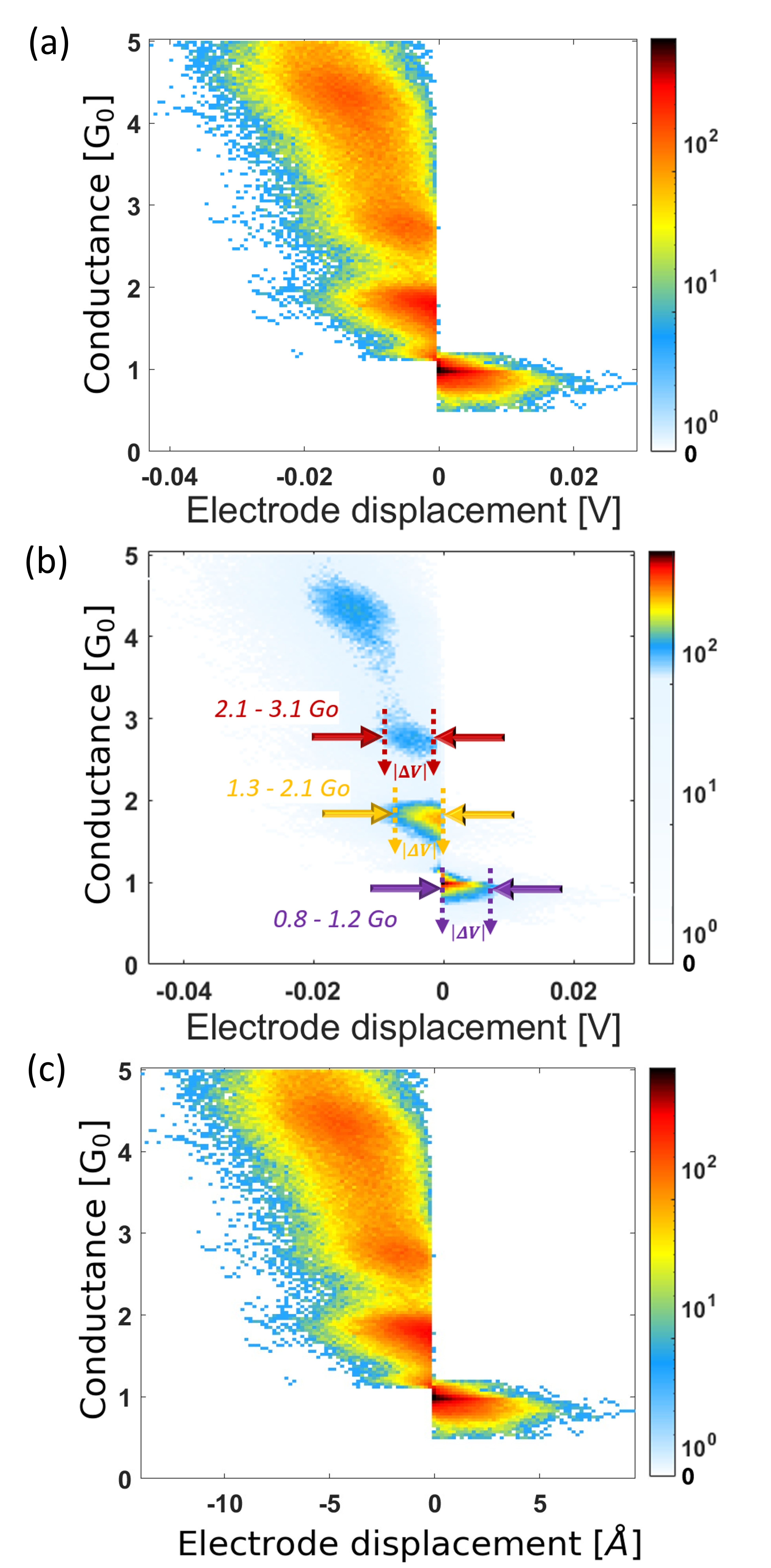}
    \caption{(a) Density plot of gold rupture traces under room conditions, where relative displacement is not calibrated and units are in volts. (b) A two-dimensional plot was employed for calibration, with color counts representing only the maximum and one order of magnitude less. (c)  Density plot of gold rupture traces under room conditions.}
    \label{Denspanel}
\end{figure}

However, the depicted trace provides only half of the information, as the relative distance can be expressed not only in volts but also in angstroms. To carry out the calibration and following the methodology described above, we have generated a density plot using the data from the rupture traces, ensuring that all of them are expressed in units of relative displacement in volts applied to the piezo. For the construction of the density plot, we will ensure that all traces are offset to the same point, where this is the first value less than or equal to $1G_0$ (see Figure \ref{Denspanel} panel(a)). The color bar is displayed on a logarithmic scale, spanning from 0 counts (white color) to approximately 800 counts (black color). The red/orange color was selected as the reference for the most representative area.Therefore, the areas marked in red/orange show very distinctive characteristics, and their relative distance in volts is very similar. Thus, following the previous discussion at 4.2 K, this means that these relative distances approximately measure 2.5 \text{Å}.

To highlight the fact that the first, second, and third plateaus exhibit similar relative displacements, we processed the data as shown in Figure \ref{Denspanel} (b). This panel is the same as panel (a), except that in this case, the minimum value of the colorbar has been adjusted in order to highlight the values that span nearly two orders of magnitude. This visual enhancement helps to better distinguish the initial and final bins, which are otherwise not clearly visible. As a result, we can more easily identify the specific initial and final bins that will be used. From this cloud of points, the first three bins and the last three bins are selected to calculate an average value in mV units, thereby obtaining a relative voltage difference. The arrows on this panel, along with the color code used throughout the manuscript, indicate the relative distance expressed in voltage units $(\Delta V$).  This voltage will be expressed in absolute value (see Table \ref{Tab:cal}), as if we are below zero, they will be negative (the range between 1 and 3 $G_0$ is positioned to the left, meaning in negative values, while the range between 0 and 1 $G_0$ is positioned to the right, meaning in positive values). As we have seen in panel (b), the three cyan-colored point clouds measure approximately the same distance in absolute value units $\Delta V$. As we have reiterated throughout the discussion of this manuscript, these clouds have a length of 2.5 \text{Å}. Therefore, Table \ref{Tab:cal} shows the summarized results obtained from this panel (b).

\begin{table}[h!]
\centering
\caption{The first column displays the conductance range for the cloud, the second column represents the absolute value in units of mV of the relative displacement, and the third column denotes the common value estimated for these plateaus.}
\vspace{12pt}
\centering

\begin{tabular}{|c|c|c|}
\hline
\begin{tabular}{@{}c@{}}Conductance \\range ($G_0$)\end{tabular}  & $ |\Delta V|$ (mV) &\begin{tabular}{@{}c@{}}Atomic distance \\ Au ($\text{\AA}$)$^{[27]}$\end{tabular}    \\ \hline
\textcolor{red}{2.1 - 3.1}  & \textcolor{red}{7.3}   &  \multirow{3}{*}{2.5}  \\ \cline{1-2}
\textcolor{gold}{1.3 - 2.1} & \textcolor{gold}{8.1}  & \\ \cline{1-2}
\textcolor{rep}{0.8 -1.2}   & \textcolor{rep}{7.6}   &  \\ \hline
\end{tabular}
\label{Tab:cal}
\end{table}

Therefore, by calculating the mean value and the standard deviation of the data in Table \ref{Tab:cal}, we obtain that the calibration factor is \(7.7 \pm 0.3 \, \text{mV}\), which corresponds to 2.5  \text{Å}. Thanks to this calibration factor, we can now calibrate all our traces and density plots. Panel Fig. \ref{Denspanel} (c) represents the calibrated density plot, where it can be observed that the areas from light orange to red approximately correspond to 2.5 {\text{Å}.

Once our system is calibrated, it is fundamental to create histograms of length, just as was done for the gold data at low temperatures and with the STM-BJ. In Figure \ref{histoLMCBJ}, a length histogram is shown for plateaus ranging from 0.8 to 1.2 $G_0$ conductances (panel (c)). In panel (b),  the histogram corresponding to plateaus ranging from 1.2 to 2.1 $G_0$ is presented. Finally, in panel (a), the length histogram is represented for plateaus ranging from 2.1 to 3.0 $G_0$. All these relative displacement length histograms, obtained from MCBJ measurements at room temperature, show that the vast majority of events fall within the range corresponding to the atomic diameter of gold, which is indicated by a vertical green line. 

\begin{figure}[h!]
 \centering
 \includegraphics[width=0.45\textwidth]{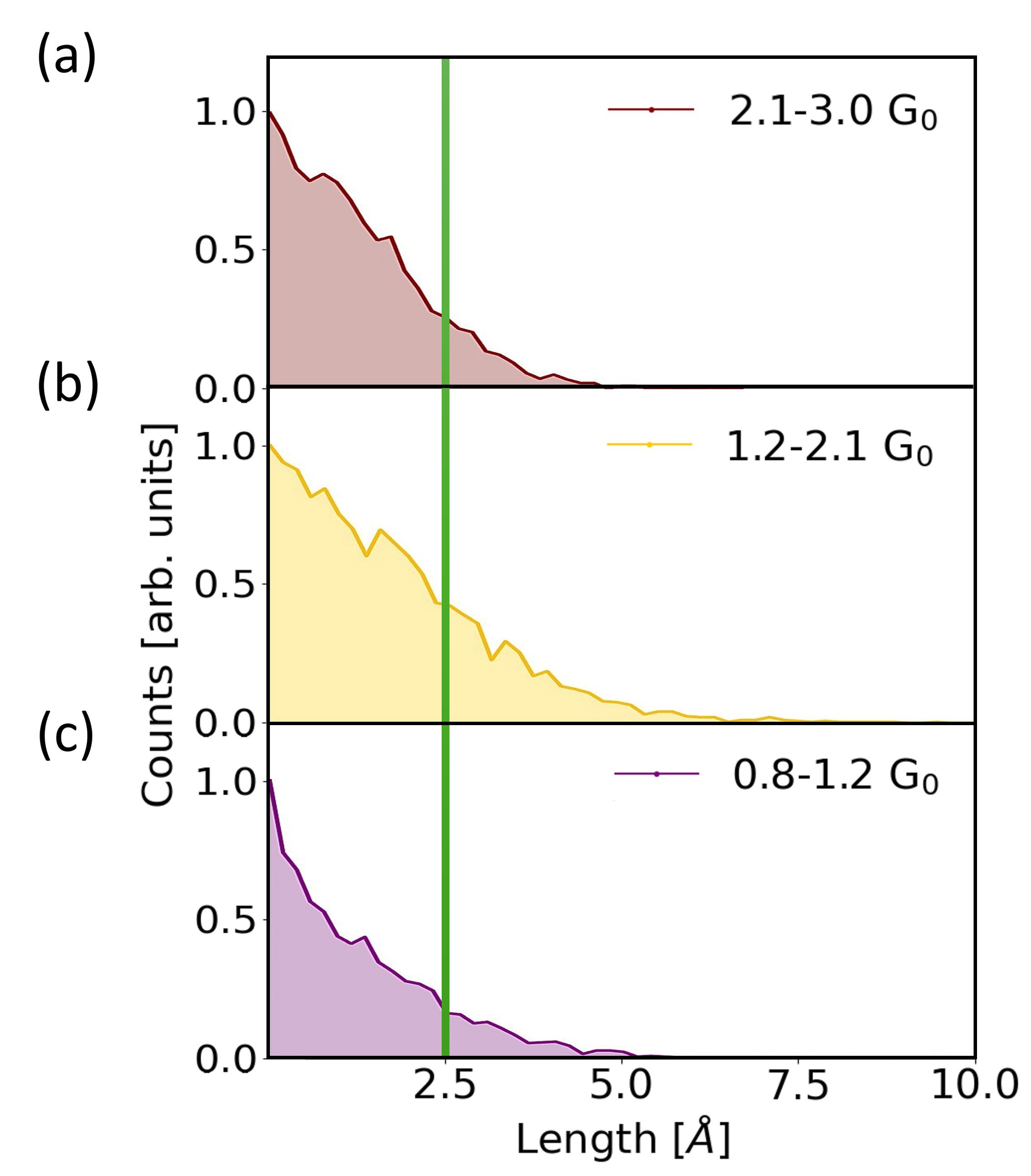}
    \caption{Histogram of lengths panel (a) 2.1-3.0 $G_0$ (red),  histogram of lengths paenl (b) 1.2-2.1 $G_0$ (yellow), histogram of lengths panel (c) 0.8-1.2 $G_0$(purple). Vertical green line denotes the atomic gold diameter.}
    \label{histoLMCBJ}
\end{figure}

Once all calibrations have been completed, we can represent the relative displacement of the trace in Å as shown in Figure \ref{TraceCal}. An inset is presented to emphasize three plateaus that correspond to atomic structures. As a curiosity  the plateau corresponding to the structure of three atoms thick is longer in distance than the first and second plateaus. 

\begin{figure}[h!]
 \centering
 \includegraphics[width=0.49\textwidth]{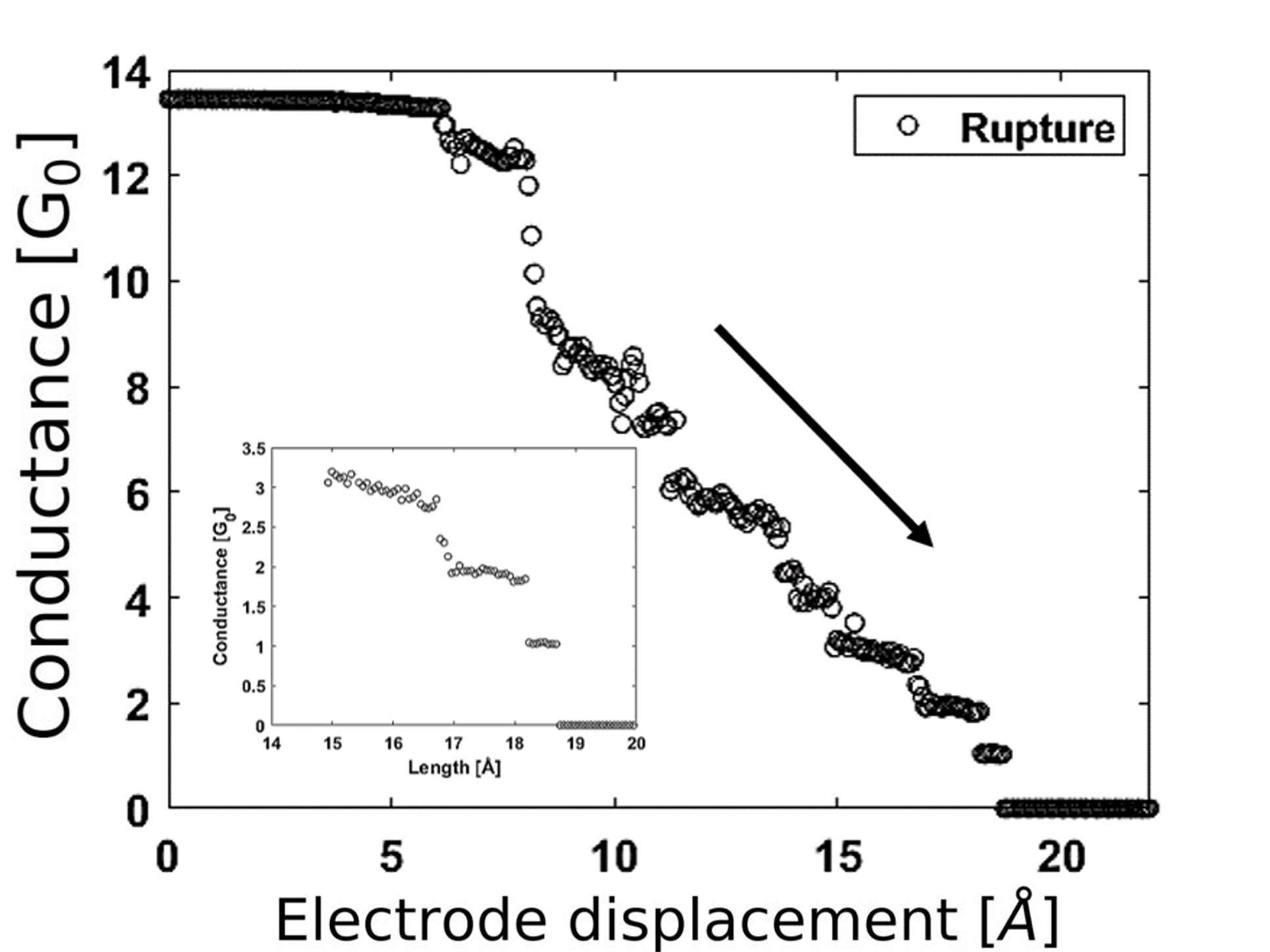}
    \caption{Calibrated trace of conductances, where the relative length electrodes is expressed in Å. The Inset is a zoom-in  order to observe the shape of the plateaus in the range of 3 to 1 $G_0$.}
    \label{TraceCal}
\end{figure}

In our manuscript, we evaluate how the cross-section of the electrode decreases when it is stretched, which reveals how the electrode tapers under tension, both at low temperatures and at room conditions. To do this, we consider that gold (Au) at 4.2~K forms atomic chains (one atom thick) and diatomic chains (two atoms thick) of variable lengths, in contrast to room condtions that only form atomic contacts with diffent thincknees. 

Since our method is based on evaluating how the conductance slope varies with distance, long plateaus affect the overall slope of the analyzed trace. Therefore, to avoid these plateaus influencing our statistics, we have decided to adopt a precautionary approach: using structures that are above the 5 $ G_0$ threshold.   This method is based on obtaining the decay slopes, once the relative displacements are calibrated, from 12$G_0$ to 5$G_0$.

Figure \ref{Ilustrace} shows an illustration of two different electrodes while being stretched. The image on the left shows the shape of its cross-section, presenting steeper slopes due to a more abrupt variation, as seen in the red curve in the central graph. On the other hand, the illustration on the right shows a smaller cross-section, where moving by an atomic distance disconnects fewer atoms are disconnected, resulting in a lower slope compared to the previous case (see blue curve). In other words, depending on the cross-section, slopes that tend to zero will indicate a smaller cross-section, while slopes that tend to infinity will suggest a larger cross-section.

\begin{figure}[h!]
 \centering
 \includegraphics[width=0.49\textwidth]{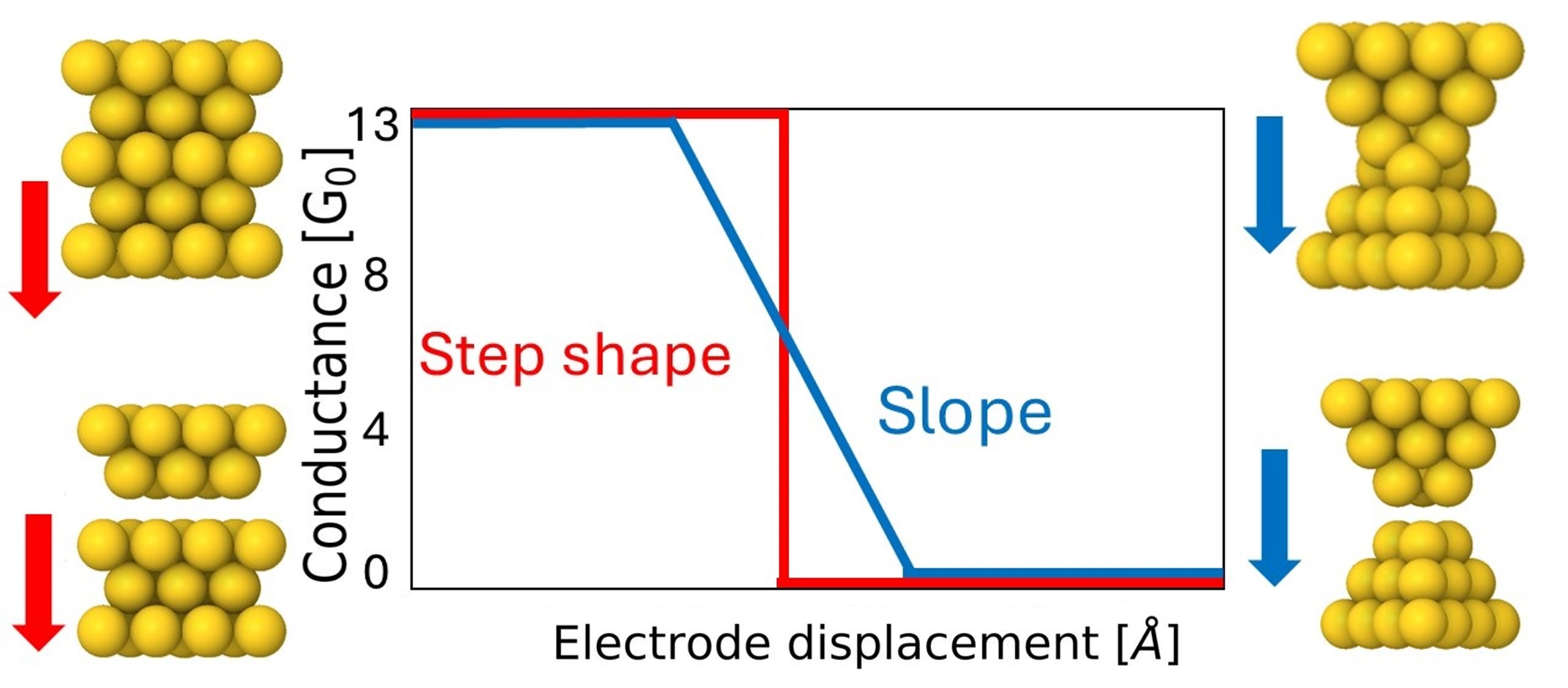}
    \caption{Illustration of two electrodes with different cross-sections. The central plot shows how the slope decreases for the different cross-sections, with the wider one represented in red and the smaller slope in blue.}
    \label{Ilustrace}
\end{figure}

Once all the slopes are fitted (see Figure \ref{slopeAu1trace} in SI), we create a histogram of the slopes in units of $G_{0}/\text{Å}$, where all slopes are computed as absolute values, as shown in Figure \ref{HistoSlope}.

\begin{figure}[h!]
 \centering
 \includegraphics[width=0.49\textwidth]{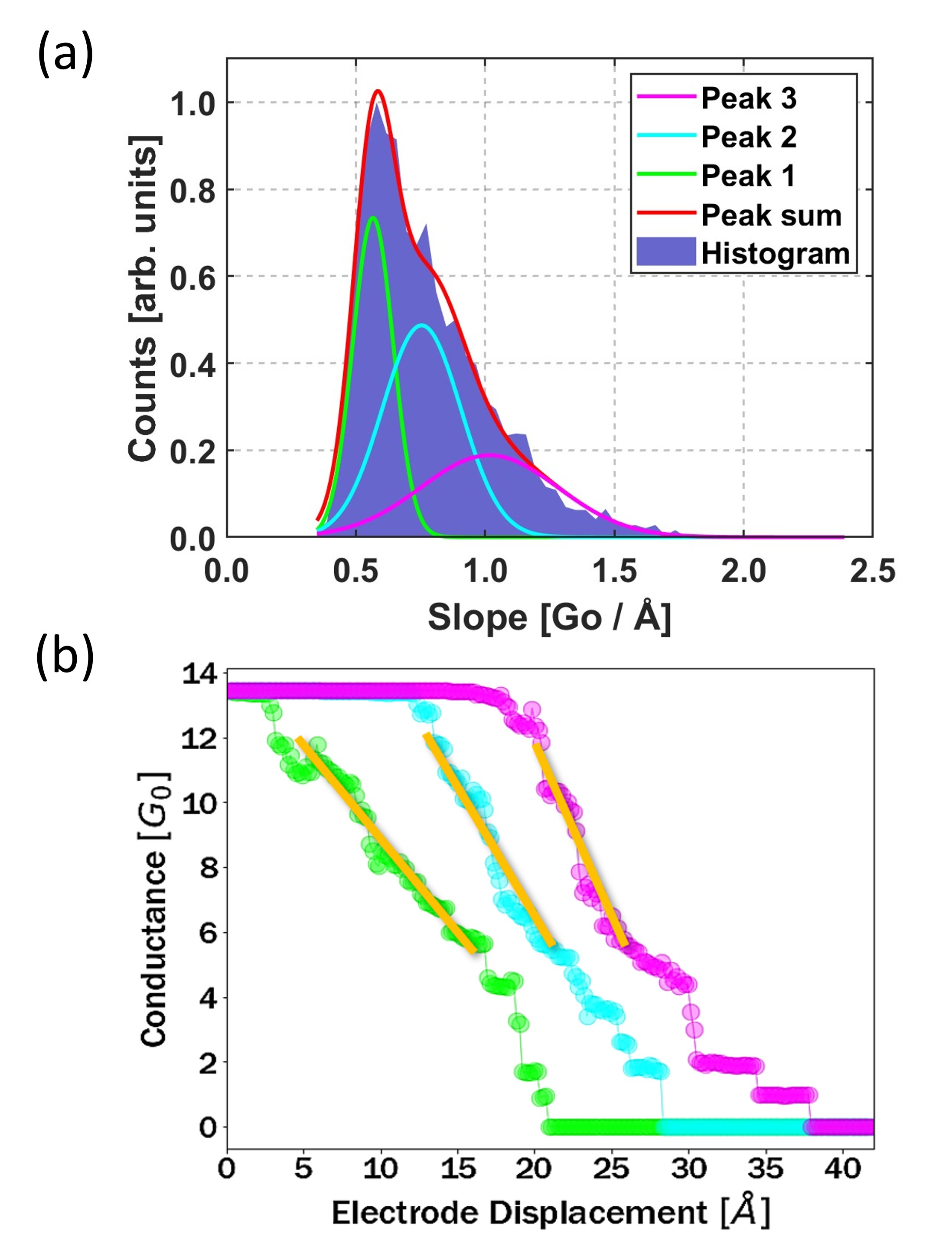}
    \caption{(a) Normalized histogram of conductance decay slopes \(G_0/\text{\AA}\). The histogram represents the distribution of slope values obtained from rupture traces, fitted with Gaussian components (colored curves) corresponding to different characteristic slopes. (b) Representative rupture traces of conductance as a function of relative displacement, where each color corresponds to a specific slope distribution in the histogram above, illustrating different levels of electrode sharpness and structural evolution.}
    \label{HistoSlope}
\end{figure}

It is clear that all the slopes are negative; however, to bring clarity to the histogram, we will use the absolute values.
Figure \ref{HistoSlope} shows a normalized histogram of the 5000 slopes analyzed.
Based on the previous results, we identified three distinct conductance decay slopes located at 0.58, 0.82, and 1.06 in units of \( G_0/\text{Å} \). To gain insight into sharpness, we suggest thinking in terms of atomic distances. For example, if we assume that these are the three slopes we have and that the electrodes are moved by 2.5 Å, this would imply a conductance reduction of approximately 1.45, 2.05, and 2.62 
$G_0$   for the three cases. In other words, in the first case, the cross-section is reduced by at least one atom, in the second case by at least two atoms, and in the last case by at least three.

These slopes provide insights into the atomic-scale evolution of electrode sharpness during the rupture process. Specifically, the most typical slopes in terms of atomic distances are 1.45, 2.05, and 2.62 \( G_0/\text{Å} \), which can be interpreted as different sharpness levels of the electrodes. The sharpest electrodes exhibit lower slopes, corresponding to the disconnection of a single atomic pathway per atomic displacement, whereas broader electrodes exhibit higher slopes, suggesting a loss of multiple atomic connections per unit of displacement. (see Sup. Inf. section Gaussian Fitting for more details).

In Figure. \ref{HistoSlope}, the upper panel shows the histogram of slopes, which is fitted by the addition of three Gaussians.The green curve corresponds to the most probable case (slope is 0.5 \( G_0/\text{Å} \) ; the blue and pink curves represent the second and third most probable cases, respectively. In Figure. \ref{HistoSlope}, the bottom panel displays three representative traces with different slopes, whose values correspond to the centers of the Gaussian distributions in the upper panel, following the same color code. Specifically, the green trace corresponds to a slope of 0.5 \( G_0/\text{Å} \), the blue trace to 0.8 \( G_0/\text{Å} \), and the pink trace to 1 \( G_0/\text{Å} \). In summary, the representation of the traces shows a transition from the sharpest (left) to the least sharp (right).
\begin{figure}[h!]
 \centering
 \includegraphics[width=0.49\textwidth]{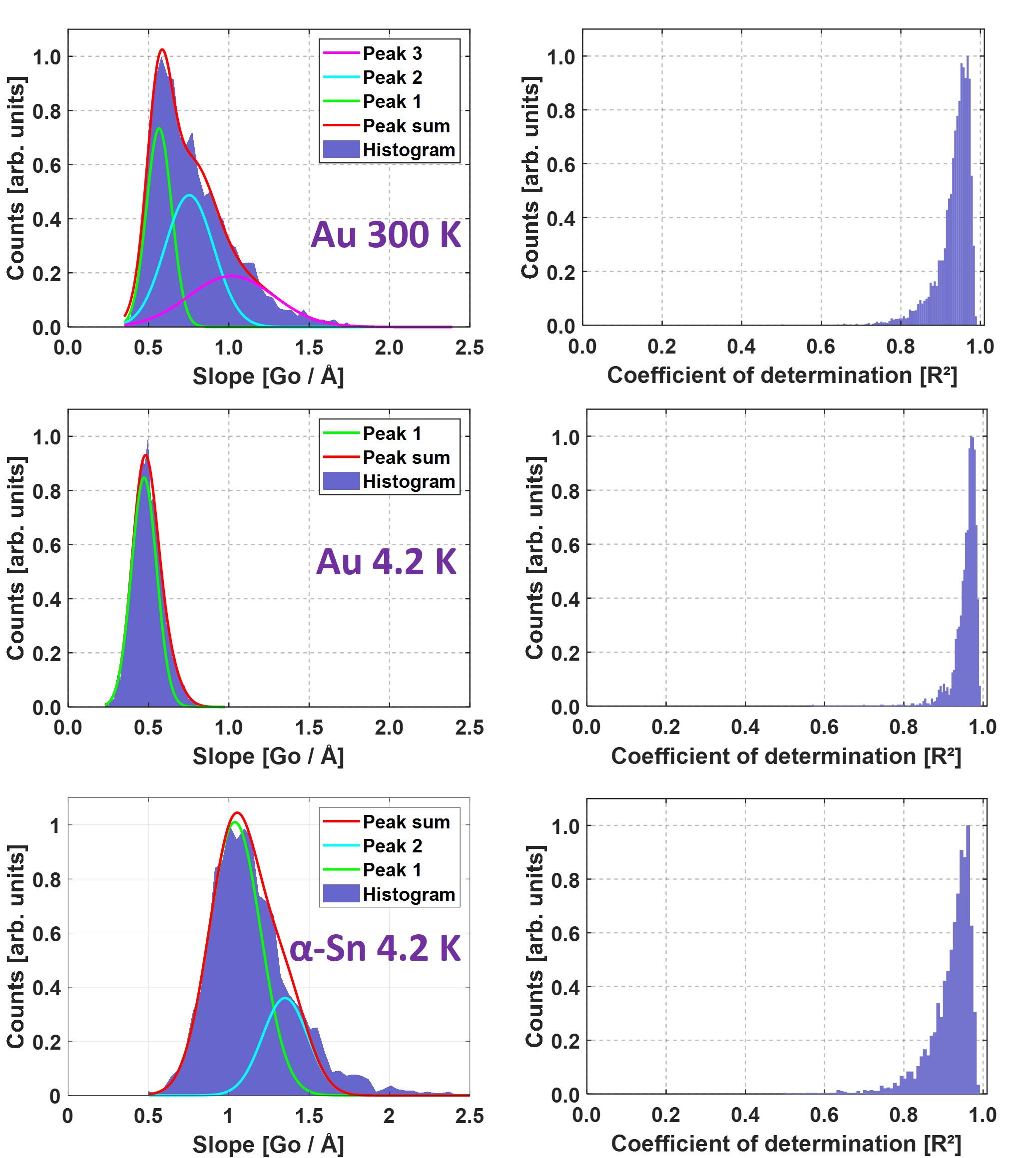}
    \caption{Conductance slopes and regression for different materials and temperatures. Left panels show normalized histograms of slopes (\( G_0/\text{\r{A}} \)) from 5000 rupture traces for gold (\( \text{Au} \)) at 300 K (top), 4.2 K (middle), and alpha-tin (\( \text{\(\alpha\)-Sn} \))  at 4.2 K (bottom). Gaussian fits highlight distinct slope distributions. Right panels display the coefficient of determination (\( R^2 \)) for the linear fits, confirming the robustness of the slope extraction.}
    \label{fig:slopes_all}
\end{figure}

However, we do not only aim to extend this new methodology to assess how sharp our electrodes are as a function of stretching, but also to apply it at different temperatures, to systems such as Au at 4.2 K and other materials like alpha-tin ($\alpha$-Sn). This material is particularly interesting because, among other characteristics, the conductance of a single atom is approximately 2.2 $G_0$ \cite{SNYanson}.

Figure \ref{fig:slopes_all} shows, in the left column, the slope histograms for atomic gold at room temperature, at 4.2 K, and for $\alpha$-Sn at low temperatures. The right column displays the histograms of the coefficient of determination $R^{2}$. As observed in all these histograms, the data are closely clustered around $R^{2} = 0.95$.

Comparing the histograms for gold at 4.2 K, we see that there is only one peak corresponding to the value 0.47 \( G_0/\text{Å} \), which is in complete agreement with the most probable value for gold at 300 K. Against all expectations, this suggests that gold remains very sharp at low temperatures when stretched, whereas at room temperature, the sharpening occurs more frequently.

On the other hand,  $\alpha$-Sn exhibits two distinct Gaussian distributions, one centered at 1.03 and the other at 1.37. Considering only the first Gaussian and knowing that a $\alpha$-Sn atom has an interatomic distance of 2.9 Å \cite{RSn}, we estimate that if we displace by this distance, the conductance reduction upon stretching decreases by approximately 2.9 $G_0$. This implies a reduction of at least one Sn atom, which, as mentioned previously, has a conductance of 2.2 $G_0$. For the other remaining case, if we displace by at least one atomic distance, we would decrease the conductance by approximately 4 $G_0$, that is, at least two Sn atoms.

\section{Conclusions}

In this work, we have experimentally identified the existence of three-atom-thick gold structures in atomic contacts and their typical triangular stacking arrangement, expanding the previous understanding limited to one- and two-atom-thick chains. The combination of Break-Junction (BJ) experimental techniques, including STM-BJ and MCBJ, along with molecular dynamics simulations and \textit{ab initio} electronic transport calculations, has enabled us to validate these findings and propose a novel and robust calibration methodology for BJ systems.

The proposed method allows for the conversion of piezoelectric displacement into absolute distances using the characteristic length of gold contacts at different thickness states. Additionally, we have developed a technique to assess the sharpness of electrodes during elongation, providing key insights into the geometry and structural evolution of atomic contacts.

The obtained results have direct applications in the precise calibration of Break-Junction systems, even under ambient conditions, which are typically challenging to calibrate due to the dependence on work functions and the necessity of characterizing the metal under these conditions. This represents a significant advancement in the characterization of metallic contacts at the atomic scale. This methodology could also be extended to other materials and studies in molecular electronics, enabling a better understanding of transport properties in metallic and molecular nanostructures, thereby providing insights into their geometry and mechanical properties at the atomic scale.

These findings not only contribute to the advancement of BJ system calibration but also allow us to explore new structures. Even for experiments under room-temperature conditions, such as those involving silver (Ag) and copper (Cu), our methodology can be used for calibration. Moreover, the slope method has provided new insights into how materials sharpen or remain stable during the stretching process, and how many atoms are disconnected during this process. We propose that this method could be extended to molecular electronics, where the slope could reveal how the decoration of molecules on the wire affects its geometry and sharpening.

\section{Acknowledgement}
The authors acknowledge financial support from the Generalitat Valenciana through CIDEXG/2022/45, CIGRIS/2021/159, and PROMETEO/2021/017, as well as from the Spanish Government through MFA/2022/045, PID2019-109539-GB-C41, PID2022-141712NB-C22, and PID2023-146660OB-I00. This work is part of the Advanced Materials programme and has been supported by MCIN with funding from the European Union NextGenerationEU (PRTR-C17.I1) and the Generalitat Valenciana (MFA/2022/045). Additionally, the authors acknowledges funding from MICIU/AEI/10.13039/501100011033 (PID2023-146660OB-I00) and the European Regional Development Fund (ERDF/EU) under project PID2023-146660OB-I00. Theoretical modeling was carried out at the Applied Physics Department of the University of Alicante, utilizing the Matcon Cluster, led by Prof. M.J. Caturla.

\bibliography{references.bib}
\section{Supplemental Material}
\subsection{Slope fitting}
Fig. \ref{slopeAu1trace} shows a conductance trace (red dots) and the linear regression fit for the trace in the range from 12 to 5 $G_0$.

\begin{figure}[h!]
 \centering
    \includegraphics[width=0.49\textwidth]{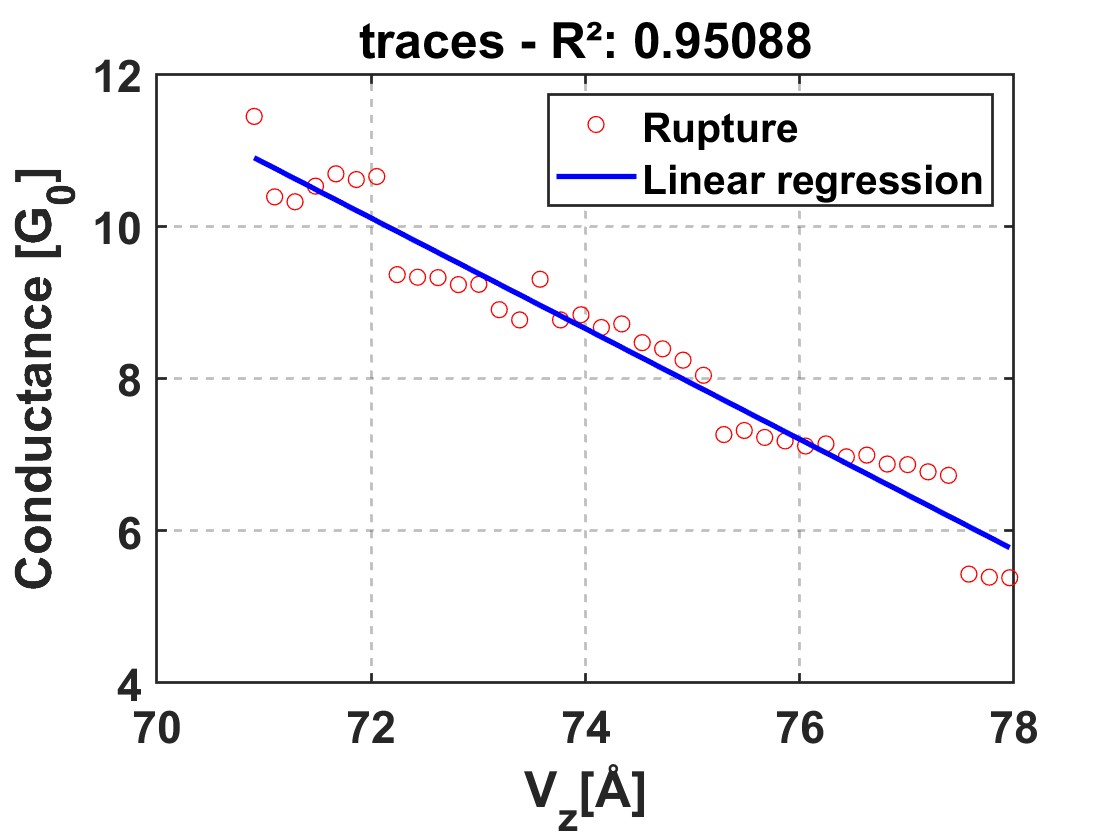}
    \caption{The red dots correspond to the gold trace obtained at 300 K in MCBJ, while the blue line represents the linear fit. As shown in the upper part of the plot, the coefficient of determination $R^2$ is 0.95.}
    \label{slopeAu1trace}
\end{figure}

\subsection{Gaussian fitting}
 To gain further insight into the types of slopes present in our data, we fitted three Gaussians in the upper panel of Figure \ref{HistoSlope}. The mathematical expression for the fitting function is given by:

\begin{equation}
f(x) = \sum_{i=1}^{N} \frac{A_i}{w_i \sqrt{\pi / 2}} e^{-\frac{(x - x_{c,i})^2}{2w_i^2}}
\end{equation}

where \(i\) corresponds to the index of the Gaussians. The fitting function is a sum of \(N\) Gaussians, where each Gaussian is defined by its amplitude \(A_i\), center \(x_{c,i}\), and width \(w_i\). \(A_i\) controls the area under the curve of the \(i\)-th Gaussian, \(x_{c,i}\) represents the center or mean, and \(w_i\) determines the width or standard deviation of the curve. The normalization ensures that the total area under the curve is proportional to \(A_i\), while \(w_i\) regulates the spread of the curve.

In this study, the Gaussians were fitted using least-squares fitting, and Table II shows the parameters obtained from this fitting, which yielded an \(R^2\) value of 0.9907. The high \(R^2\) value indicates that the model accurately fits the experimental data, specifically referring to the fit of the sum of the Gaussians.

\begin{table}[h!]

\caption{Gaussian parameters for each peak. The first column refers to the peak number, the second column to the amplitude (\(A_i\)), the third column to the center of the Gaussian (\(x_{c,i}\)), the fourth column to the width (\(w_i\)), and the fifth and sixth columns show the height and the Full Width at Half Maximum (FWHM), respectively.}

\centering
\begin{tabular}{|c|c|c|c|c|}
\hline
Peaks & \(A_i\) & \(x_{c,i}\) & \(w_i\)  & FWHM \\ \hline
\rowcolor[HTML] {99FF99}
1     &  0.7340   & 0.5797       & 0.1108    & 0.2609 \\ \hline
\rowcolor[HTML]{99FFFF} 
2     &  0.4869   & 0.8214       & 0.2151    & 0.5065 \\ \hline
\rowcolor[HTML]{FF99FF}
3     &  0.1882   & 1.0590       & 0.3754      & 0.8840 \\ \hline

\end{tabular}
\label{histofit}
\end{table}

\end{document}